\documentstyle[prb,preprint,aps]{revtex}

\begin{document}

\draft

\title{Dielectric nonlinearity of relaxor ferroelectric ceramics at 
low ac drives}

\author{Xiao-Wen Zhang and Xi Liu}
\address{Department of Materials Science and Engineering, 
Tsinghua University, Beijing 100084, People's Republic of China}
\author{Zhi-Rong Liu and Bing-Lin Gu}
\address{Department of Physics, Tsinghua University, Beijing 100084, 
People's Republic of China}
\address{Center of Materials Research, Tsinghua University, Beijing 
100084, People's Republic of China}

\maketitle

\begin{abstract}
Dielectric nonlinear response of 
(PbMg$_{1/3}$Nb$_{2/3}$O$_3$)$_{0.9}$(PbTiO$_3$)$_{0.1}$ 
(0.9PMN-0.1PT) relaxor ceramics was investigated under 
different ac drive voltages. It was observed that: 
(i) the dielectric permittivity is independent on ac field amplitude 
at high temperatures; (ii) with increasing ac drive, the permittivity 
maximum increases, and the temperature of the 
maximum shifts to lower temperature; (iii) the nonlinear effect is 
weakened when the measurement frequency increases. The influences of increasing 
ac drive were found to be similar to that of decreasing frequency. 
It is believed that the dielectric nonlinearities of relaxors at low 
drives can be explained by the phase transition theory of ergodic space 
shrinking in succession. A Monte Carlo simulation 
was performed on the flips of micro 
polarizations at low ac drives to verify the theory.
\end{abstract}


\vspace{2mm}


\section{Introduction}

Since Pb(Mg$_{1/3}$Nb$_{2/3})$O$_3$ (PMN) was first synthesized by Smolenski 
{\it et al.} in the late 1950s,\cite{1} there has been a series of 
relaxor ferroelectrics (relaxors) 
with complex perovskite structure whose dielectric and ferroelectric 
properties are rather different from that of normal ferroelectrics. 
For the relaxor ferroelectrics, the dielectric permittivity is unusually high, 
the sintering temperature is rather low, and the temperature coefficient 
of capacitance is quite small due to the diffuse phase transition (DPT), 
which lead to the successful application as 
Multi-layered Capacitors (MLC).\cite{2} In addition, 
the field-induced piezoelectric effect of relaxors is strong, 
and the pulse echo response of transducer can 
be controlled by bias voltage. So the relaxors are competent in the range 
of actuators, medical diagnostic transducers, etc..\cite{3,4} 
Recently, the observation of the highly excellent electromechanical 
properties in some single crystal of relaxors 
(for example, the PMN-PT solid solution) has brought great interest to 
the research, development and 
application of this kind of materials.\cite{5}

In actual applications, the components and device made 
of relaxors usually work under dc bias or ac drive voltages. 
So the performances of the material under external field always cause great 
interests. In recent years, some works focus on the 
dielectric nonlinear response under various ac drive voltages. 
Apart from the strong application background, these works can also 
provide some clues of the 
polarization mechanism of relaxors in theory. Although several 
possible models have been proposed, the nature of the dielectric 
response of relaxors, especially PMN, keeps unclear.\cite{6} 
Experiments such as high-resolution transmission electron microscopy 
(HTEM) have confirmed a main feature of PMN in structure: 
a great amount of nanoscaled ordered microregions are embedded randomly 
in the disordered matrix.\cite{7,8} The ordered microregions 
are probably nonstoichiometric.\cite{8} The assumption that 
the ordered microregions are just the centers of polar 
microregions presented by the superparaelectric model\cite{9} need further 
testification of experiments, but the theoretical calculation based 
on the assumption can explain the particularly large 
permittivility of PMN.\cite{10} 

For normal ferroelectrics such as BaTiO$_3$, the nature of dielectric 
nonlinearity at different ac drives has been universally 
accepted, i. e., the nonlinearity is caused by the movement of 
domain walls among the ferroelectric domains with 
different polarization directions. 
For relaxor ferroelectrics, opposite with normal ferroelectrics, no 
macro phase transition takes place, and no ferroelectric domain 
with micrometric dimensions appears. Under the usual conditions, 
the nanoscaled polar microregions in relaxors do not grow into 
ferroelectric domains even if the temperature is much lower than 
$T_{max}$ (the temperature where the permittivity reaches a maximum). 
Therefore, it is greatly interesting to investigate the connections  
between the dielectric nonlinear response and the polar microregions.

The low frequency dielectric properties of PMN at different 
ac drives were first reported by Bokov et al. in the early 1960s.\cite{11} 
But there have been little works reported on this subject 
until 1990s. In the recent decade, with the widely 
application of PMN-type relaxors and the increasing interests on 
the polarization mechanism of relaxors, the research of 
nonlinearity was emphasized again. Experiments showed that 
$T_{max}$ shifts to lower temperature with increasing ac drive amplitude 
for PMN.\cite{12,13} And the case of PLZT with relaxor behavior 
is similar.\cite{14,15} 
The curves of permittivity at various 
drives are similar to the frequency dispersion in relaxors. Glazounov 
et al.\cite{12} presented that the nonlinear behavior was controlled 
by the domain wall motion rather than the reorientation of polar 
clusters (i. e., superparaelectric approach\cite{9}). 
However, they did not explain neither the type of 
domains nor the process of domain wall motion. 
Colla et al.\cite{16} investigated the experiments of PMN-PT 
single crystals and pointed out that the nonlinearity 
mechanism is related to the drive amplitude: a glass-like dynamics 
of the polarization freezing process is dominant at low drives; 
at intermediate drives, the movements and reconstruction of 
the boundaries of polar nanodomains take place; at higher 
drives, the interactions between polar regions cause the 
formation of normal micron-sized domains and the movement of 
domain walls.

In this paper, the nonlinear dielectric response of 0.9PMN-0.1PT 
ceramics at various ac drive amplitudes and 
frequencies was studied. The experimental 
results were qualitatively explained by the phase transition 
theory of ergodic space shrinking in succession, and a 
Monte Carlo simulation was conducted to verify the theory.

\section{Experimental procedure}

0.9PMN-0.1PT powder was prepared by the columnbite precursor 
method.\cite{17} The starting materials are analytically pure PbO, 
Nb$_2$O$_5$, TiO$_2$, and (MgCO$_2$)$_4$.Mg(OH)$_2$.6H$_2$O. 
MgNb$_2$O$_6$ was synthesized by (MgCO$_2$)$_4$.Mg(OH)$_2$.6H$_2$O and 
Nb$_2$O$_5$ at 1000 $^\circ$C. MgNb$_2$O$_6$, PbO, and TiO$_2$ 
powders were mixed and calcined at 870 $^\circ$C for 2 h. 
Then the PMN-PT 
powders obtained were pressed into pellets 
($\phi$10$\times$1--2 mm) at 100 MPa, and sintered in the 
PbO-rich atmosphere  
for 2 h at 1200 $^\circ$C. The specimen were 
analyzed by the X-ray diffraction technique on a 
diffractometer (Science D/max-RA) using CuK$_{\alpha}$ raciation, 
and a pure perovskite structure was confirmed. Finally, 
the specimen were polished to 0.4 mm, and plated by silver. 
The dielectric permittivity was measured using a 
HP4284 LCR meter over the frequency range 1--100 kHz at 
a heating rate 3 K/min. The amplitude of the ac measuring 
field is 0.05, 0.25, 0.40, and 0.50 kV/cm.

\section{Results and Discussions}

The usual amplitude of the ac signal used to measure the dielectric 
permittivity of relaxors is 0.01 kV/cm.\cite{12} The 
results obtained correspond to the slope of the hysteresis loop, 
$\partial P/\partial E$, at the starting point. Because the amplitude 
is small enough to fall in the linear-response region, 
$\partial P/\partial E$ is constant, which represents the dielectric 
permittivity $\varepsilon'$, i. e., the $\varepsilon'$ value 
is independent on the amplitude of external field. However, with 
the increasing of ac field amplitude, 
the nonlinear terms can not be ignored.

Fig.1--Fig.3 show the change in the dielectric permittivity 
measured at various amplitudes when the ac field frequency is 
1 kHz, 10 kHz and 100 kHz, respectively. From the figures one can 
list the following features: (1) when the measuring frequency 
is fixed, the dielectric permittivity keeps constant for various 
ac drives at high temperatures, while it increases with 
increasing the ac amplitude at low temperatures; 
(2) with increasing the amplitude, the dielectric permittivity 
maximum, $\varepsilon_m'$, increases and shifts to lower 
temperature; (3) the diffusion behavior is more evident at 
larger ac amplitude; (4) the dielectric nonlinear effect is 
weakened at lower frequencies. Fig. 4 demonstrates the dielectric 
permittivity at various frequencies when the amplitude is fixed 
as 0.05 kV/cm. 
(The curves for $E=$0.25, 0.4 and 0.5 kV/cm are omitted since 
they are similar to Fig.4.) It is noted that increasing amplitude has the 
same effects on the permittivity as decreasing 
frequency.

It can be seen from the results above that the effect of 
ac field amplitude on the permittivity maximum $\varepsilon_m'$ 
is obvious. The relation between $\varepsilon_m'$ and amplitude 
is depicted in Fig. 5. A linear law is found in the range 
of amplitude and frequency under study, which is consistent 
with the results of single crystal.\cite{16} 
With increasing the frequency, the effect of amplitude on $\varepsilon_m'$ 
is weakened. Extrapolate the curves in Fig. 5 to the zero field, 
we can obtain the permittivity maximum without nonlinear effect. 
Table I gives the 
variation of $\varepsilon_m'$ when the amplitude increases 
from 0.05 kV/cm to 0.5 kV/cm. It shows that $\varepsilon_m'$ 
increases by 7.4\% at 1 kHz (which is the frequency in usual measurements). 
This means that the measuring result of permittivity 
is affected by the weak-field 
nonlinearity. As a result, 
the thickness and the applied voltage of specimen, i. e., 
the field strength, should be specified to avoid the 
influence of nonlinear effect, so that the results in different 
experiments are comparable.

It was mentioned in Sec. I that there were different explains on the 
nonlinear effect at weak fields.\cite{12,16} Glazounov et al.\cite{12} 
denied the mechanism of 
reorientation of polar clusters, but they have ignored the interactions 
between polar clusters. Colla et al.\cite{16} 
presented that a glass-like dynamics of the polarization freezing 
process dominates at low drives. However, the connections between the 
dynamic process and the nonlinear effect were not explained.

We proposed that the dielectric nonlinearity of relaxor ferroelectrics 
at low drives, as well as the frequency dispersion, can 
be explained in the theory of the phase transition of ergodic 
space shrinking in succession.\cite{18} The TEM dark-field image of 
0.9PMN-0.1PT proved that there are a great amount of nanoscaled 
ordered microregions embedded in the disordered matrix.\cite{19} 
At a certain temperature, the homogeneous crystal structure of the 
ordered microregions causes the cooperative displacement of B-site 
cations along one of eight $\langle 111\rangle$-equivalent directions. 
When the temperature is high enough, the thermal energy, $k_BT$, is much 
larger than the energy barriers between different directions, which 
results that the probabilities of displacement along eight directions 
are equal, i. e., the system is ergodic. Thus the ordered microregions are 
unpolar. However, the environments of the microregions along different 
directions are not identical, and the spherical symmetry of ordered 
microregions breaks down. Since the potential wells of different 
$\langle 111\rangle$ directions are different, the B-site ions tend to 
stay along the direction with the lowest well for more time in 
the thermal flipping 
process when the temperature decreases to a certain value. Thus the 
ordered microregions transform into the polar microregions, and 
dipole behavior appears. Due to the cooperative displacement, the 
ions in the same microregion flip as a whole unit under external drives, 
so the dielectric 
permittivity of relaxor ferroelectrics is extraordinarily high. The polar 
microregions are random the magnitude and direction of polarization. 
Under zero field, $\sum p_i=0$, while $\sum p_i^2\not= 0$. When the 
temperature is much higher than the freezing temperature, the relaxation 
times of polar microregions are much shorter than the observation  
time. All the polar microregions flip dynamically with ac drives, so the 
sum of polarizations, $P$, is proportional to the external field, i. e., 
the dielectric permittivity, $\varepsilon=\partial P/\partial E$, is 
independent on the frequency and the amplitude of external field. In this 
temperature range, the deviation of micro-polarization direction from that of external 
field by thermal fluctuation is weakened with decreasing 
temperature, and then the permittivity increases with decreasing 
temperature. This corresponds to the high temperature regions in 
Fig.1--Fig.4, where no frequency dispersion and nonlinearity is found. 
The polar microregions can be regarded as independent dipoles. 
When the temperature further decreases, the electrostatic 
interactions between dipoles get more and more strong. Under the 
ac drives with a certain frequency, the flip of a dipole is 
affected by both the external field applied and the internal field 
generated by other dipoles. Some dipoles 
cannot keep up with the switching of the measuring field, 
and become ``slow dipoles". Some are even frozen along 
a certain direction, and become ``frozen dipoles". Thus the phase 
space with ergodicity is shrinking in succession. When the frequency 
increases, the time scale of dipole flipping is shortened. More dipoles 
cannot reach the equilibrium states in the observation 
time, i.e., the proportion of slow dipoles and frozen dipoles 
increases. Slow dipoles and frozen dipoles give no or little contribution 
to the flipping polarization. So the dielectric permittivity decreases, 
which is the frequency dispersion in relaxors. When the ac field 
amplitude increases, the driving force on dipoles is enhanced. Slow 
dipoles and frozen dipoles are forced to flip faster and give more 
contribution to the flipping polarization. 
The proportion of slow dipoles and frozen dipoles decreases, and 
the dielectric permittivity increases, which is the nonlinear 
effects in relaxors. This is the origin of the special 
dielectric properties in relaxors (frequency dispersion, 
nonlinearity, etc.).

It should be emphasized that the external field discussed above 
is in the range of weak drives (less than 0.6 kV/cm according to 
Ref. 16). Only at weak drives can nanoscaled dipoles exist, and 
the long-range interactions between dipoles dominate in the 
dynamic process. If the external field increases, motions and 
reconstruction of the boundaries of polar nanodomains would take 
place. Under the strong fields, through interacting with 
the disordered matrix surrounding, polar nanodomains will switch, 
coalesce, and grow into the conventional micron-sized ferroelectric 
domains as that in normal ferroelectrics.\cite{16,20} Then the model 
discussed above is not applicable to describe the 
polarization dynamics.

In order to better understand the dielectric nonlinearity in 
relaxors and verify the  
model above, a Monte Carlo simulation is conducted in the next
section to investigate 
the dynamic flipping process of the polar microregions .

\section{Monte Carlo simulation}

Gui et al.\cite{21} have used the Monte Carlo method to simulate 
the dynamics of freezing process in relaxor ferroelectrics. 
In the theory framework of Ref. 21, the polar microregions are modeled 
as point dipoles. The interaction between two dipoles with 
moment $\stackrel{\rightarrow}{\mu}_i$ and 
$\stackrel{\rightarrow}{\mu}_j$ is expressed as 
\begin{eqnarray}
J_{ij} = J_{ji} & = & -\stackrel{\rightarrow}{\mu_{j}} \cdot
         \frac{1}{ 4\pi \varepsilon_{0}}  
         \left(
         \frac{3 \stackrel{\rightarrow}{\mu_i}
         \cdot \hat{r}_{ij} }    {r^3}
         \hat{r}_{ij}-\frac{\stackrel{\rightarrow}{\mu_i}}{r^3}
         \right) \nonumber \\
       & = & -\frac{1}{4\pi \varepsilon_{0}}\cdot
            \frac{3\cos\varphi_{i}\cos\varphi_{j}-\cos
         \phi}{r^{3}}\mu_{i}\mu_{j},
\end{eqnarray}
where $\hat{r}_{ij}$ is the unit vector between the two dipoles. 
$r$ is the distance between dipoles. $\varphi_i$ ($\varphi_j$) is 
the angle between $\stackrel{\rightarrow}{\mu_{i}}$ 
($\stackrel{\rightarrow}{\mu_{j}}$) and $\hat{r}_{ij}$. $\phi$ is 
the angle between $\stackrel{\rightarrow}{\mu_{i}}$ and 
$\stackrel{\rightarrow}{\mu_{j}}$. The Hamiltonian of relaxors at 
dc bias is obtained as 
\begin{equation}
H=\frac{1}{2} \sum\limits_{i\neq j}J_{ij} 
    - E\sum\limits_{i}\mu_{i}\cos\theta_i,
\end{equation}
where $E$ is the dc field strength. $\theta_i$ is the angle 
between $\stackrel{\rightarrow}{E}$ and 
$\stackrel{\rightarrow}{\mu}_i$.

The effective interaction energy, $\stackrel{\sim}{J_{ij}}$, 
is introduced as
\begin{equation}
\stackrel{\sim}{J_{ij}}\sigma_i\sigma_j
=J_{ij}\mu_i\mu_j/2, 
\end{equation}
where $\sigma_{i}=\pm1$ is the projection of 
$\stackrel{\rightarrow}{\mu_i}$ on the direction of the external 
field. Eq. (2) can be rewritten as 
\begin{equation}
H=-\sum\limits_{i\not{= }j}\stackrel{\sim}{J_{ij}}\sigma_{i}
\sigma_{j}-E \overline {\mu}\sum\limits_{i}{\frac{|\mu_{i}\cos
\theta_{i}| }{\overline{\mu}}}\sigma_{i},
\end{equation}
where $\overline {\mu}$ is the maximal projection of dipole moments 
on the external field.

Ref. 21 used Eq. (4) to study the dielectric origins 
of relaxor ferroelectrics under dc external field. It is 
unsuitable for the case of ac field. However, the polarization mechanism 
should be similar in both ac and dc fields. Therefore, an ac field 
term is introduced as 
\begin{equation}
E(t)=E\cos\left(2\pi\frac{t}{t_L}\right),
\end{equation}
where $t_L$ is the period of the ac field, which corresponds to the 
frequency. Thus a Hamiltonian similar to Eq. (4) is 
obtained:
\begin{equation}
H=-\sum\limits_{i\not{= }j}\stackrel{\sim}{J_{ij}}\sigma_{i}
\sigma_{j}-E(t) \overline {\mu}\sum\limits_{i}{\frac{|\mu_{i}\cos
\theta_{i}| }{\overline{\mu}}}\sigma_{i}.
\end{equation}
A Gaussian distribution is assumed for $\stackrel{\sim}{J_{ij}}$, 
i. e., 
\begin{equation}
P(\stackrel{\sim}{J_{ij}})\propto 
      exp[-\frac{\stackrel{\sim}{J_{ij}}^{2}}{2(\Delta J)^2}],
\end{equation}
where $\Delta J$ is the distribution width.

There are N dipoles in the system. (N=$16\times 16\times 16$) 
The flipping probability of the $i$th dipole is 
defined as 
\begin{equation}
W=\frac{1}{e^{\delta H/k_B T}+1},
\end{equation}
where $\delta H$ is the change of energy when the dipole 
flip from $\sigma_i$ to -$\sigma_i$. The details of simulation process 
can be found in Ref. 21. Then the ac permittivity can be obtained as
\begin{equation}
\chi=\frac{1}{E}\cdot\frac{1}{t_{obs}} \int\limits_{t_0}^{t_0+t_{obs}}
p(t)\exp\left(i2\pi\frac{t}{t_L}\right)dt,
\end{equation}
where $p(t)$ is the average polarization.

Fig. 6 shows the temperature dependence of dielectric permittivity 
at various ac amplitudes when the frequency is fixed as 
$t_L$=5 MCS/dipole. It can be seen that the diffusion behavior 
is enhanced with 
increasing amplitude. The simulation results are consistent with 
experiments in main features, which verifies the 
polarization mechanism described by the above model. 
It is noted that the curve of $E$=3.0$\Delta J/\overline{\mu}$ 
(which corresponds to a stronger 
field) lies below other curves at high temperatures. 
In this case, maybe the corresponding field is too strong and cause 
the growth of dipoles, so the weak-field model is not applicable.

\section{Conclusions}

The dielectric nonlinear response of 0.9PMN-0.1PT ceramics 
was revealed over the field range 0.05--0.50 kV/cm. When the 
measurement frequency is fixed, the dielectric permittivity is 
invariant with field amplitude at high temperatures. At low temperatures, 
the permittivity maximum, $\varepsilon_m'$, increases and shifts 
to lower temperatures with increasing amplitude. 
A linear law between $\varepsilon_m'$ and the amplitude was observed at 
all frequencies and amplitudes in the experiment. 
The nonlinearity is weakened at 
higher frequencies. The effects of increasing amplitude are similar to 
that of decreasing frequency. It was proposed that the 
nonlinearity of relaxors at low drives can be explained in the 
theory of the phase transition of ergodic space shrinking in 
succession. A Monte Carlo simulation was conducted to investigate 
the dynamic flips of polar microregions at low drives and 
verify the proposition.


This work was supported by the Chinese National Science Foundation
(Grant NO. 59995520).

\begin{table}
\caption{
 The variation of dielectric permittivity maximum 
when the amplitude increases from 0.05 kV/cm to 0.5 kV/cm.
\label{Table1}}
\begin{tabular}{|c|c|}
frequency (kHz) & change of $\varepsilon_m$ (\%)\\
\tableline\tableline
1 & 7.4 \\ \tableline
10 & 6.4 \\ \tableline
100 & 2.0\\
\end{tabular}
\end{table}

\begin{figure}[tbp]
\caption{ The dielectric permittivity of 0.9PMN-0.1PT as functions 
of temperature at various amplitudes when the frequency is fixed 
as 1 kHz. }
\end{figure}

\begin{figure}[tbp]
\caption{ The dielectric permittivity of 0.9PMN-0.1PT as functions 
of temperature at various amplitudes when the frequency is fixed 
as 10 kHz. }
\end{figure}

\begin{figure}[tbp]
\caption{ The dielectric permittivity of 0.9PMN-0.1PT as functions 
of temperature at various amplitudes when the frequency is fixed 
as 100 kHz. }
\end{figure}

\begin{figure}[tbp]
\caption{ The dielectric permittivity of 0.9PMN-0.1PT as functions 
of temperature at various frequency when the amplitude is fixed 
as 0.05 kV/cm. }
\end{figure}

\begin{figure}[tbp]
\caption{ The dielectric permittivity maximum of 0.9PMN-0.1PT 
at various amplitudes and frequencies. }
\end{figure}

\begin{figure}[tbp]
\caption{ The simulation results of dielectric permittivity 
at various amplitudes when $t_L$=5MCS/dipole. The temperature is 
measured in unit of $\Delta J/k_B$, and amplitude in unit of 
$\Delta J/\overline{\mu}$. }
\end{figure}

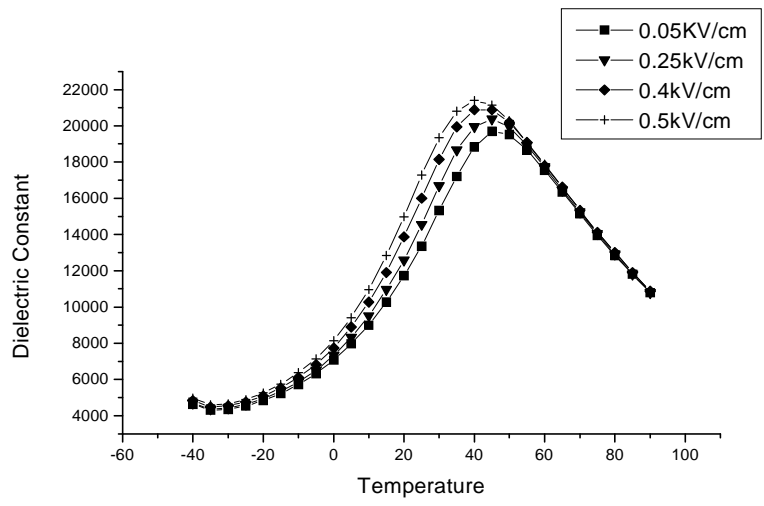
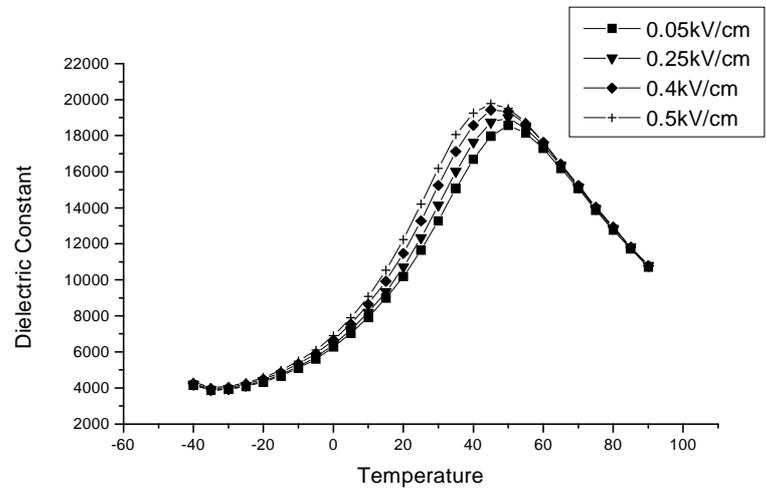
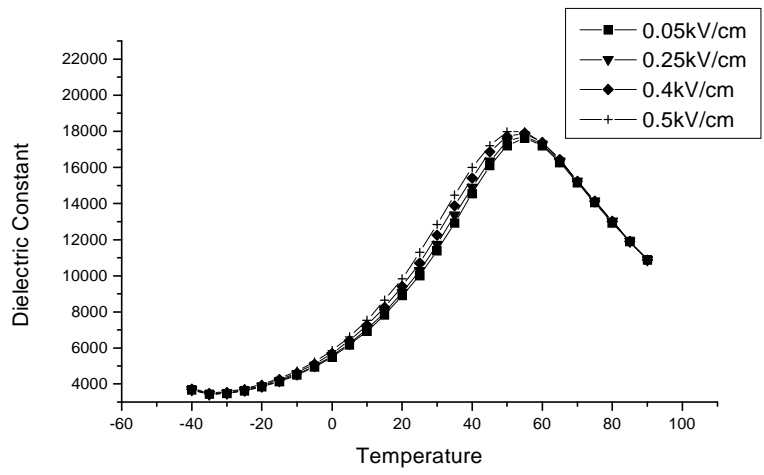
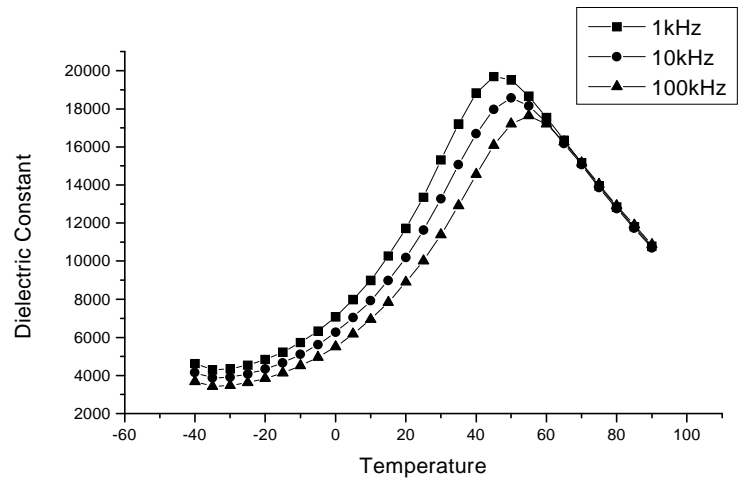
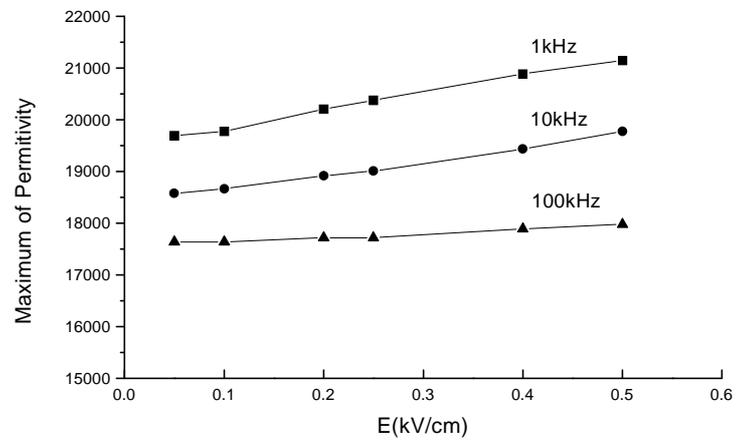
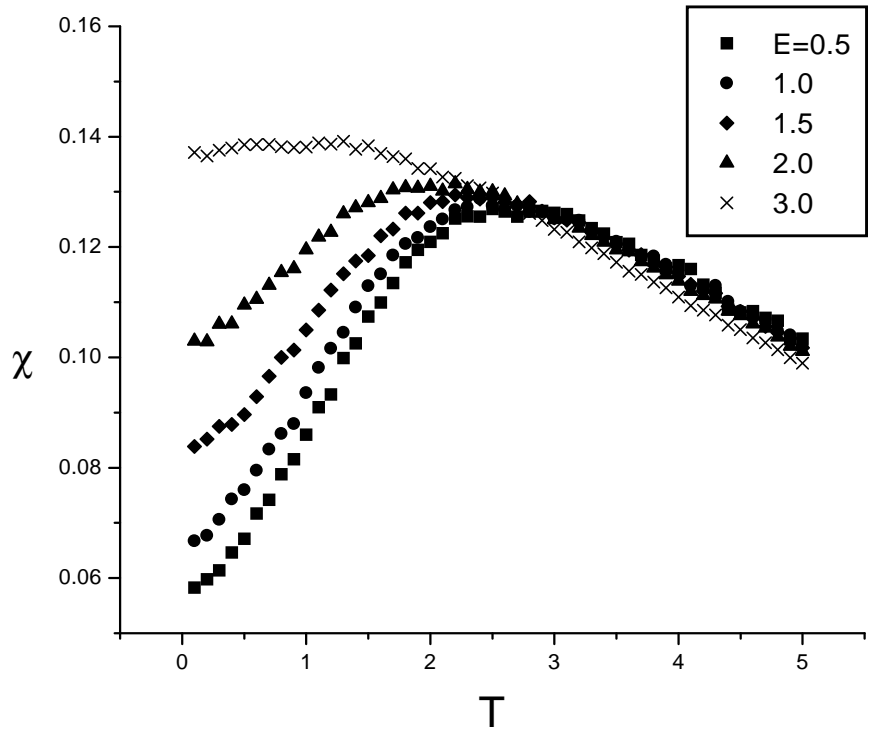
\end{document}